\documentclass[10pt,superscriptaddress,twocolumn,amsmath,amssymb,aps,prx]{revtex4}
\usepackage{times}
\usepackage{amsfonts}
\usepackage{mathrsfs}
\usepackage{graphicx}% Include figure files
\usepackage{dcolumn}% Align table columns on decimal point
\usepackage{bm}% bold math
\usepackage{color}

\usepackage[colorlinks,bookmarks=false,citecolor=blue,linkcolor=red,urlcolor=blue]{hyperref}
\usepackage{mhchem}

\def\be{\begin{equation}}       \def\ee{\end{equation}}
\def\bea{\begin{eqnarray}}      \def\eea{\end{eqnarray}}

\begin{document}
\title{Predicting Diamond-like Co-based Chalcogenides as Unconventional High Temperature Superconductors}

\author{Jiangping Hu}\email{jphu@iphy.ac.cn}
\affiliation{Beijing National Laboratory for Condensed Matter Physics,
and Institute of Physics, Chinese Academy of Sciences, Beijing 100190, China}
\affiliation{Kavli Institute of Theoretical Sciences, University of Chinese Academy of Sciences,
Beijing, 100190, China}
\affiliation{Collaborative Innovation Center of Quantum Matter,
Beijing, China}

\author{Yuhao Gu}
\affiliation{Beijing National Laboratory for Molecular Sciences, State Key Laboratory of Rare Earth Materials Chemistry and Applications, Institute of Theoretical and Computational Chemistry, College of Chemistry and Molecular Engineering, Peking University, 100871 Beijing, China}
\affiliation{Beijing National Laboratory for Condensed Matter Physics,
and Institute of Physics, Chinese Academy of Sciences, Beijing 100190, China}

\author{Congcong Le}
\affiliation{Beijing National Laboratory for Condensed Matter Physics,
and Institute of Physics, Chinese Academy of Sciences, Beijing 100190, China}
\affiliation{Kavli Institute of Theoretical Sciences, University of Chinese Academy of Sciences,
Beijing, 100190, China}

\begin{abstract}
We predict Co-based chalcogenides with a diamond-like structure  can host unconventional high temperature superconductivity (high-$T_c$). The essential electronic physics in these materials  stems from the Co layers  with each layer being formed by vertex-shared CoA$_4$ (A=S,Se,Te) tetrahedra  complexes,  a material genome proposed  recently by us to  host  potential unconventional high-$T_c$ close to a $d^7$ filling configuration in  3d transition metal compounds.  We calculate the magnetic ground states of  different transition metal compounds with this structure.  It is found that (Mn,Fe,Co)-based compounds all have a G-type antiferromagnetic(AFM) insulating ground state while Ni-based compounds are paramagnetic metal.  The AFM  interaction is the largest in the Co-based compounds as the three $t_{2g}$ orbitals all strongly participate in AFM superexchange  interactions.  The abrupt quenching of the magnetism from the Co to Ni-based compounds is very similar to those from Fe to Co-based pnictides in which a C-type AFM state appears in the Fe-based ones  but vanishes in  the Co-based ones. This behavior can be considered as an electronic signature of the high-$T_c$ gene.  Upon doping, as we predicted before, this family of  Co-based compounds favor a strong d-wave pairing superconducting state.
\end{abstract}


\maketitle

\section{Introduction}

To solve the mystery of unconventional high-$T_c$ superconductivity, a necessary step is to understand  why  cuprates\cite{Bednorz1986}and iron-based superconductors\cite{Kamihara2008-jacs} are  two special types of materials to host high-$T_c$ while other materials, in particular, many other transition metal compounds that share similar structural, magnetic and electronic properties,  do not  become high-$T_c$.
A correct understanding should also be able to  guide us to search for new high-$T_c$ superconductors, in particular,  those based on other 3d transition metal compounds.

 Recently,  based on the mechanism that superconductivity is induced by the AFM superexchange interaction,  we  have suggested that  there is an electronic gene that  separates the two high-$T_c$ families  from other correlated electronic materials. Those d-orbitals that make the strongest in-plane d-p couplings in the cation-anion complexes  in both high-$T_c$ families  are isolated near Fermi surface energy\cite{Hu-tbp,Hu-genes, Hu-s-wave}. This property makes the effective AFM superexchange interactions to  maximize their contribution to superconducting pairing. The electronic gene  can only be realized in very limited special cases\cite{Hu-genes}.  It requires a perfect collaboration between  local building blocks,  global lattice structures, as well as  specific electron filling configurations in the d-shell of transition metal atoms.
 In the cuprates, the gene is realized because the  d$_{x^2-y^2}$  $e_g$ orbital can be isolated near Fermi energy in  a  two dimensional Cu-O square lattice formed by corner-shared CuO$_6$  octahedra( or CuO$_4$ square planar)  with a $d^9$ filling configuration of  Cu$^{2+}$.  In iron-based superconductors,  we have shown that  the gene condition is satisfied because two $t_{2g}$ d$_{xy}$ -type orbitals are isolated near a $d^6$ filling configuration of Fe$^{2+}$ in a Fe(Se/As)  two dimensional square lattice formed by edge-shared Fe(Se/As)$_4$ tetrahedra\cite{Hu-tbp,Hu-genes,Hu2012-s4}.

 Following the above  idea, as the d-orbitals with the strongest d-p couplings  in a cation-anion complex gain energy and move up in the energy spectra of the d-orbitals,  it is natural for us to ask first whether we can realize the gene condition in  the  $d^7$  and $d^8$ filling configurations.  Up to now, we have predicted that the $d^7$ gene  condition can be realized   in a two dimensional hexagonal  layer formed by edge-shared trigonal biprymidal complexes\cite{Hu-tbp} or in a two dimensional square lattice formed by the conner-shared tetrahedra\cite{Hu2017ASuperconductors},  and the $d^8$ gene condition exists in a two dimensional square lattice formed by Ni-based mix-anion octahedra\cite{Le2017ASuperconductors}.  Unfortunately, all these proposals have not been materialized.

 In this paper, we identify a family of  Co-based Chalcogenides with a diamond-like lattice structure, for example,  CuInCo$_2$(S, Se, Te)$_4$ with a stannite  or  primitive-mixed CuAu(PMCA)  structure, in which the Co two-dimensional  layer  is formed exactly by corner shared tetrahedra. The electronic gene is realized in the $d^7$ filling configuration of Co$^{2+}$ as we have proposed in ref.\cite{Hu2017ASuperconductors}.  The successful synthesizations of the type of  transition metal compounds  were reported in the past\cite{Delgado2016StructuralCuNi2InTe4,Gallardo2016SYNTHESIS4}. However, very few measurements have been made to study their electronic properties. Here we calculate their electronic and magnetic properties under the formula CuInM$_2$(S, Se, Te)$_4$ with M = Mn, Fe, Co and Ni. In all these materials, the electronic physics near Fermi energy stems from the M(S,Se,Te)$_2$ layer and is attributed to the d-orbitals of the transition metal elements. In the Co case,  the three $t_{2g}$ orbitals are near degenerate and close to half-filling, and make the dominating contribution near Fermi energy.  The (Mn,Fe,Co)-based compounds all have a checkerboard (G-type) AFM insulating ground state while Ni-based compounds are paramagnetic metal.  The AFM exchange  interaction is the largest in the Co-based compounds. This magnetic trend from the Co to Ni based compounds is very similar to those from Fe to Co-based pnictides in which  a C-type AFM state appears in the Fe-based ones  but vanishes in  the Co-based ones\cite{Zeng2017MagnetismPnictides}.  This result indicates that the Co-based parental compounds are  multi-orbital Mott insulators. However, upon electron doping, the magnetic long-range order can quickly be diminished and possible d-wave superconductivity can arise as we predicted in ref.\cite{Hu2017ASuperconductors,Li2017}. Experimental results appear to be consistent with our calculation results\cite{Gallardo2016SYNTHESIS4}. The new materials closely resemble both cuprates and iron-based superconductors, and can bridge the gap between their electronic properties. We believe that this family of materials can display a variety of novel electronic phases and  can be a fertile new ground to study strongly correlated electronic physics.

\section{Diamond-like transition metal chalcogenides}
The diamond-like quaternary chalcogenides with formula I-II$_2$-III-VI$_4$ and I$_2$-II-IV-VI$_4$ can be considered as  the derivatives of zinc-blende chalcogenides by sequential cation cross-substitution\cite{Chen2009ElectronicCompounds,Liang2017AnalysisStructures}. There are a  variety of possible compositions with I = Cu,
Ag; II=Zn, Cd, Mn, Fe, Co; III=Al, Ga, In; IV=Si, Ge, Sn; VI
= S, Se, Te\cite{Schafer1974TetrahedralCu2IIIVS4Se4,Delgado2008CrystalDiffraction,Delgado2014SynthesisAgFe2GaTe4,Delgado2015CrystalDiffraction,Delgado2016Crystal4,Delgado2016StructuralCuNi2InTe4,Delgado2018Synthesis4,Guo2009SynthesisCells,Chen2009ElectronicCompounds,Liang2017AnalysisStructures,Liang2017Mid-InfraredRelationship}.  In the past, a great attention was paid to study  Zn/Cd compounds for their semiconducting properties due to potential applications  for  photovoltaics\cite{Guo2009SynthesisCells}, non-linear optics\cite{Liang2017AnalysisStructures,Liang2017Mid-InfraredRelationship} and so on.  However, the transition metal compounds have not been well studied.

\begin{figure}
\centerline{\includegraphics[width=0.48\textwidth]{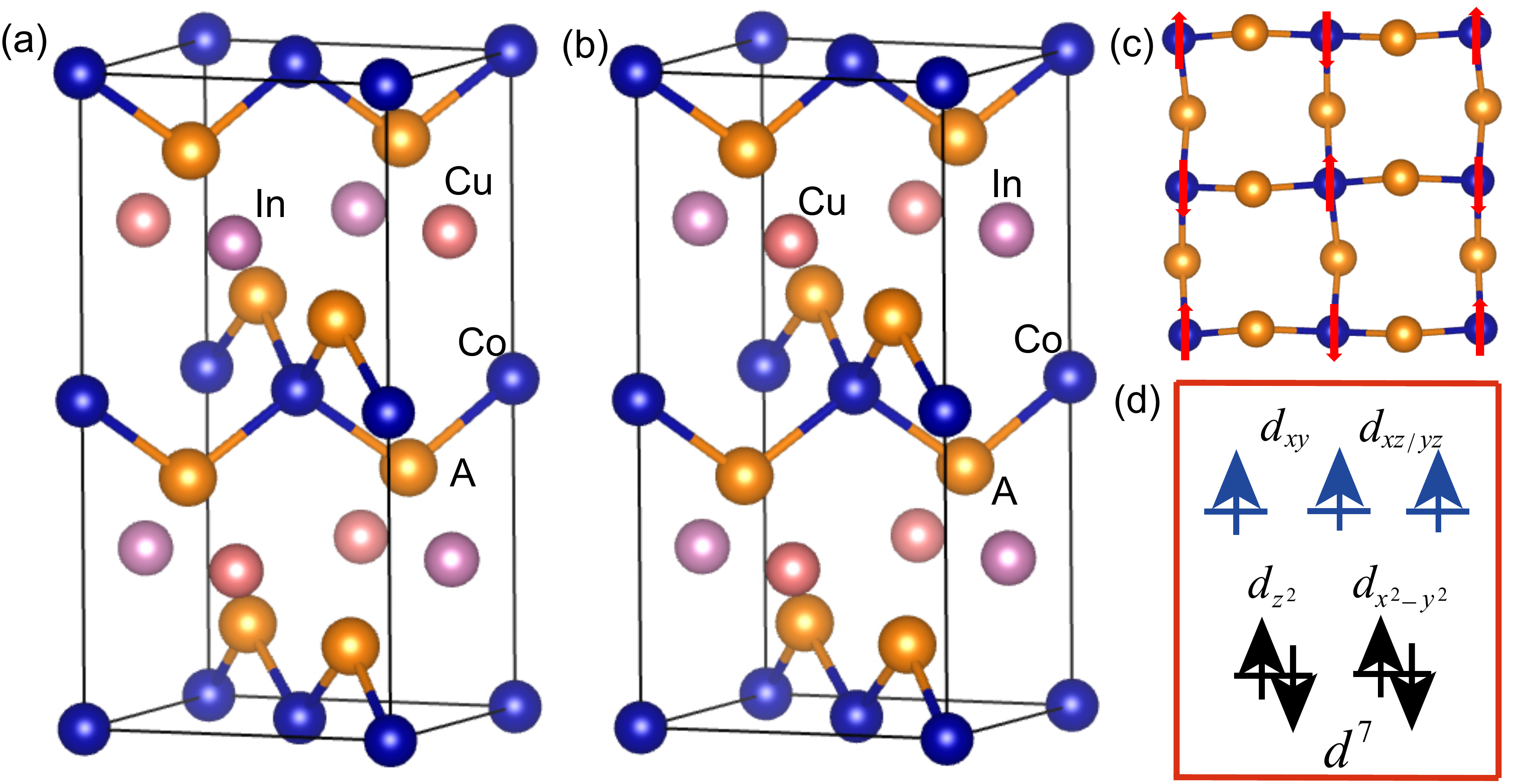}} \caption{(color online) (a) and (b) are the CuInCo$_2$A$_4$ (A=S, Se, Te) Stannite and PMCA crystal structures. (c) show the checkerboard AFM state. (b) Crystal energy splitting configurations and the required electron filling configuration.
 \label{lattice} }
\end{figure}

Here we consider the I-II$_2$-III-VI$_4$ quaternary chalcogenides with II=Mn, Fe, Co and Ni, namely, transition metal elements.  There are three different possible structures for this formula, kesterite, stannite  and PMCA\cite{Chen2009ElectronicCompounds}.  We focus on the last two structures, stannite and PMCA because in these two structures, the transition metal  chalcogenide  tetrahedra form a square lattice through corner-sharing as shown in Fig.\ref{lattice} (a) and (b).

 It is obvious that in both stannite and PMCA structures, the electronic physics  is carried out by the  partially filled 3d-shell of the transition metal atoms.  A simple picture is that by changing II=Zn to II=transition metal elements in these structures, the bands of the  d-orbitals   are filled  in the semiconductor gap. Therefore, both stannite and PMCA structures support almost identical  electronic physics.  As the stannite-type structure has been reported experimentally\cite{Delgado2008CrystalDiffraction,Delgado2014SynthesisAgFe2GaTe4,Delgado2015CrystalDiffraction,Delgado2016Crystal4,Delgado2016StructuralCuNi2InTe4,Delgado2018Synthesis4}, we present our calculation results  on the stannite-type CuInM$_2$A$_4$(M=Mn, Fe, Co, Ni, A=S,Se,Te)  in the following.

 Fig.\ref{lattice} (a) and (b) show the CuInCo$_2$A$_4$ (A=S, Se, Te) stannite and PMCA crystal structures, in which each CoA$_2$ layer is constructed by vertex shared tetrahedra.  It is worthy noting that the stannite-type \ce{CuInCo2Te4} has already been synthesized\cite{Delgado2016StructuralCuNi2InTe4}. According to the Shannon ionic radii table\cite{Shannon1976RevisedActa}, the radii of cations are quite close. The cation-anion radius ratios determine the favorable coordination number (C.N.). In this case, C.N.=4 (tetrahedral coordination) is favored in \ce{CuInCo2A4} for A=S,Se,Te and the tetragonal distortion($\eta=\frac{c}{2a}$) in \ce{CuInCo2Te4} is very small .

Our DFT calculations employ the Vienna ab initio simulation package (VASP) code\cite{Kresse1996} with the projector augmented wave (PAW) method\cite{Joubert1999FromMethod}. The Perdew-Burke-Ernzerhof (PBE)\cite{Perdew1996} exchange-correlation functional was used in our calculations. The crystal structures of  were fully relaxed with a kinetic energy cutoff (ENCUT) of 600 eV for the planewaves and the $\Gamma$-centered k-mesh of $8\times8\times4$.  The energy convergence criterion is $10^{-6}$ eV and the force convergence criterion is 0.01 eV/ \AA.

The experimental and optimized lattice parameters of the CuInCo$_2$A$_4$ quaternary chalcogenides are listed in the Table.\ref{co-lattice}. The optimized crystal parameters are about 2\% smaller than the experimental values for CuInCo$_2$Te$_4$.  This is a very typical result for DFT calculations as the calculation result is for lattice at zero temperature and the electron-electron correlation effect is underestimated.

\begin{table}%The best place to locate the table environment is directly after its first reference in text
\caption{\label{CuCo2InB4 str table}%
The experimental and optimized crystal structure parameters for {CuInCo$_2$A$_4$}({A = S, Se, Te}) with the stannite-type structure (space group $I\overline{4}2m$) in the AFM state. }
%%%% new table
\begin{ruledtabular}
\begin{tabular}{cccccccccc}
System  & a(\AA) & c(\AA) & $\eta$=c/2a & Co-A(\AA) & Co-A-Co($^{\circ}$)\\
 \colrule
 {CuInCo$_2$Te$_4$},exp\cite{Delgado2016StructuralCuNi2InTe4} & 6.20 &	12.38 & 1.00 & 2.69 & 108.96 \\
{CuInCo$_2$Te$_4$},opt & 6.06 & 12.10 & 1.00 & 2.54 & 114.91 \\
{CuInCo$_2$Se$_4$},opt & 5.66 & 11.35 & 1.00 & 2.38 & 114.21  \\
{CuInCo$_2$S$_4$},opt & 5.34 & 10.80 & 1.01 & 2.25 & 114.25  \\
\end{tabular}
\end{ruledtabular}
\label{co-lattice}
\end{table}

\section{Magnetism}
In the study of magnetism of CuInM$_2$A$_4$, the GGA plus on-site repulsion $U$ method (GGA+$U$) in the formulation of Dudarev et al.\cite{Dudarev1998Electron-energy-lossStudy} is employed to describe the associated electron-electron correlation effect. The effective Hubbard $U$ ($U_{eff}$) is defined by $U_{eff}=U-J_{Hund}$ and the Hund exchange parameter $J_{Hund}$ is $0.5$eV when $U>0$ in this paper. We consider  four different magnetic states, the paramagnetic state, the  ferromagnetic(FM) state, the G-type AFM state (shown in Fig.\ref{lattice}(c) ) and the AFM$_2$ state (AFM in the a-b plane  with FM along the c axis) as the initial guess of the relaxation.

\begin{table}[!ht]%The best place to locate the table environment is directly after its first reference in text
\caption{\label{CuInM$_2$A$_4$ str table}%
The optimized crystal structure parameters for {CuInM$_2$A$_4$} (M = Mn, Fe, Co, Ni; A = S, Se, Te) with the stannite-type structure (space group $I\overline{4}2m$) from the  initial guess of the AFM state using PBE\cite{Perdew1996}. }
%%%% new table
\begin{ruledtabular}
\begin{tabular}{cccccccccc}
System  & a(\AA) & c(\AA) & $\eta$=c/2a & Co-A(\AA) & Co-A-Co($^{\circ}$)\\
 \colrule
 {CuInMn$_2$Se$_4$} & 5.84 & 11.69&1.00&2.52&110.16 \\
 {CuInFe$_2$InSe$_4$} & 5.73 & 11.49&1.00&2.42&113.37 \\
 {CuInCo$_2$InSe$_4$} & 5.66 & 11.35 & 1.00 & 2.38 & 114.21  \\
 {CuInNi$_2$InSe$_4$} &	5.62&11.19&1.00&2.34&116.38  \\
 {CuInMn$_2$Te$_4$} & 6.28&12.55&1.00&	2.71&110.04 \\
 {CuInFe$_2$Te$_4$} & 6.16&12.25&	0.99&2.60&113.74\\
 {CuInCo$_2$Te$_4$} & 6.06 & 12.10 & 1.00 & 2.54 & 114.91  \\
 {CuInNi$_2$Te$_4$} & 6.01&11.96&1.00&2.50&116.76  \\
\end{tabular}
\end{ruledtabular}
\label{lattice-mg}
\end{table}

We find that the G-type AFM state is generally favored in the Mn, Fe and Co-based compounds.   The lattice parameters at the AFM state are listed in the Table.\ref{lattice-mg}.  The ordered magnetic moments and the band gap values in the AFM state are listed in Table.\ref{moment}.

\begin{figure}
\centerline{\includegraphics[width=0.5\textwidth]{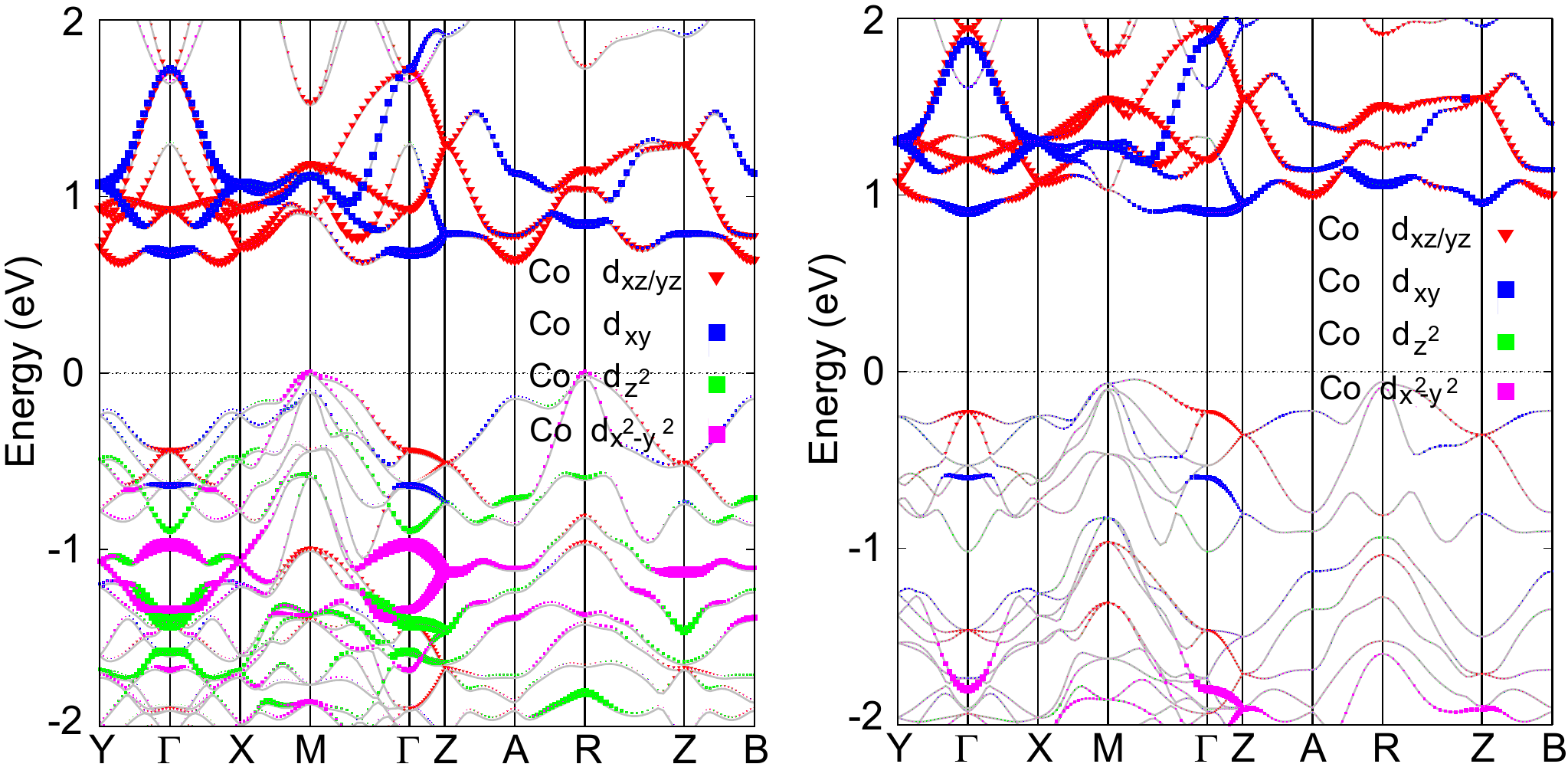}}
\caption{(color online) The band structures under the GGA+$U$ calculations for CuInCo$_2$Te$_4$: (a) $U_{eff}=2\text{eV}$; (b) $U_{eff}=4\text{eV}$.
}
\label{magnet-band}
\end{figure}

The Mn, Fe and Co atoms are, in general, in high spin states.  The magnetic moments slightly increase as $U_{eff}$ increases. Although CuInFe$_2$Te$_4$ and CuInCo$_2$Te$_4$ are metallic as shown in the Table.\ref{moment} at $U_{eff}=0$, we believe that all these compounds should be insulating because PBE often underestimates the band gap for transition metal compounds\cite{Hong2012}.  The gaps in the DFT calculations for Fe and Co-based compounds are strongly dependent on $U_{eff}$. They increase rapidly when $U_{eff}$ is switched on.   We plot the electronic band structures in the AFM state for the Co-based compounds   in Fig.\ref{magnet-band} when $U_{eff}=2.0\text{eV}, 4.0\text{eV}$.

\begin{table}
\caption{\label{moment}%
The calculated magnetic moments and  the band gaps for {CuInM$_2$A$_4$} (M = Mn, Fe, Co, Ni; A = Se, Te) using GGA+$U$ ($U_{eff}=0$ or $5\text{eV}$). }
\begin{ruledtabular}
\begin{tabular}{ccccc}
System  & Moment(\text{$\mu_B$})& Gap(eV)& Moment(\text{$\mu_B$})& Gap(eV)\\
              &  $U_{eff}=0$ & $U_{eff}=0$ &  $U_{eff}=5\text{eV}$&  $U_{eff}=5\text{eV}$\\
 \colrule
 {CuInMn$_2$Se$_4$} & 4.1 & 0.76 & 4.6 & 0.76 \\
 {CuInFe$_2$Se$_4$} & 3.0 & 0.17 & 3.6 & 0.75 \\
 {CuInCo$_2$Se$_4$} & 1.9 & 0.19 & 2.5 & 1.07  \\
 {CuInNi$_2$Se$_4$} &	0&0&1.0&0  \\
 {CuInMn$_2$Te$_4$} & 4.0 & 0.80 & 4.6 & 0.90  \\
 {CuInFe$_2$Te$_4$} & 3.0 & 0 &	3.6 & 0.92\\
 {CuInCo$_2$Te$_4$} & 1.7 & 0.02 & 2.4 & 1.04   \\
 {CuInNi$_2$Te$_4$} & 0&0&0.9&0 \\
\end{tabular}
\end{ruledtabular}

\end{table}

 In the Ni-based compounds, the magnetic states gain little energy up to $U_{eff}=2\text{eV}$, which indicates  that the Ni-based materials are paramagnetic (or very weak FM) metals.  The moment is developed fully only when $U_{eff}>4\text{eV}$. But even at $U_{eff}=5\text{eV}$, the material remains to be a metal. This abrupt quenching of the magnetic order from the Co-based to Ni-based compounds resembles the similar quenching of the magnetic order in iron-based superconductors in which  the AFM order vanishes completely when Fe is replaced by Co.

\begin{figure}
\centerline{\includegraphics[width=0.5\textwidth]{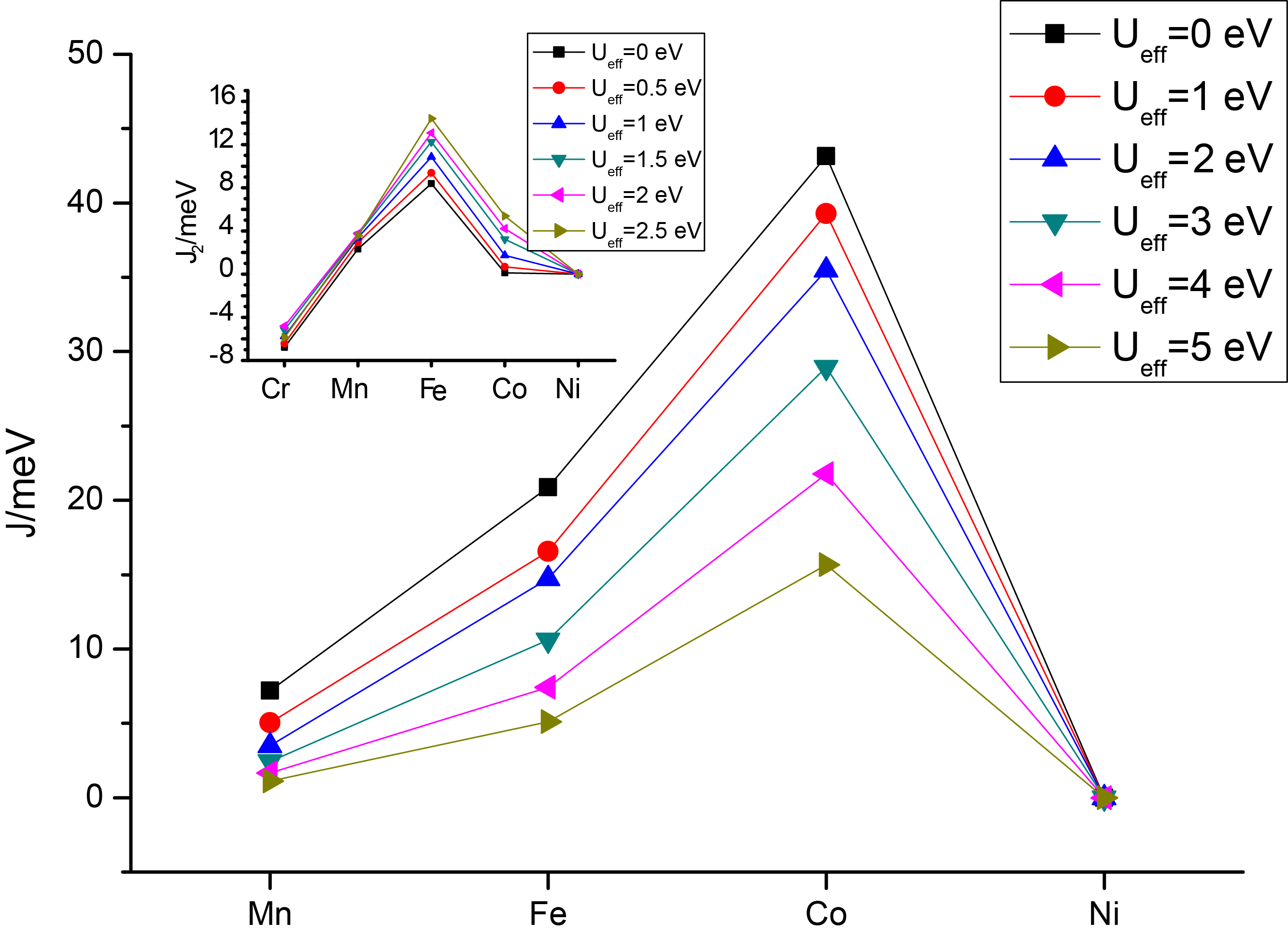}}
\caption{The J superexchange  AFM interactions  in CuInM$_2$Te$_4$  (M = Cr, Mn, Fe, Co, Ni), which are extracted from the GGA+$U$ calculations with the values $U_{eff}= (0,1,2,3,4,5) \text{eV}$. CuInNi$_2$Te$_4$ is converged to a nonmagnetic state. The inset figure is the J$_2$ AFM exchange interactions   of BaM$_2$As$_2$ (M = Cr, Mn, Fe, Co, Ni) in the 122 tetragonal iron-based superconductor structure\cite{Zeng2017MagnetismPnictides}. }
\label{J_value}
\end{figure}

 If we use an effective Heisenberg model to describe the AFM exchange interaction,  as the dominant AFM interactions are  between two nearest neighbor (NN) sites,  the model can be written as
 \begin{eqnarray}
 H=J\sum_{<ij>} \vec S_i\cdot \vec S_j+J_z\sum_{<ij>_c} \vec S_i\cdot \vec S_j
 \end{eqnarray}
 where $<ij>$ labels the in-plane nearest neighbor (NN) links and $<ij>_c$ labels the out of plane NN links. The classical energies of the above magnetic states are
 \begin{eqnarray}
& & E_{FM}=NS^2(2J+Jz)+E_0,\nonumber \\
& &E_{AFM}=NS^2(-2J-Jz)+E_0,\nonumber \\
& &E_{AFM_2}=NS^2(-2J+Jz)+E_0
\end{eqnarray}
From the calculated energies of the four different magnetic states,  we can extract the effective magnetic exchange interactions.  The interaction between the Co layers, J$_z$, is very small. It is about an order of magnitude smaller than the in-plane interaction, J.
The values of J  in CuInM$_2$Te$_4$ are plotted in Fig.\ref{J_value} as a function of M, transition metal elements.  J reaches the maximum value when M=Co. For M=Ni,  the estimation is not reliable as the moment is too small.   The results in CuInM$_2$Se$_4$ are very similar.  These qualitative results do not depend on U.  Here, the physics of J is exactly identical to  the next NN AFM interactions, J$_2$ , in iron based superconductors.  For a comparison,  we also insert a picture  in Fig.\ref{J_value} to show J$_2$ in BaM$_2$As$_2$ with M=Cr, Mn, Fe, Co\cite{Zeng2017MagnetismPnictides}.

\section{Electronic structures and Effective Hamiltonian at low energy}
In the ref.\cite{Hu2017ASuperconductors}, we have studied the CoS$_2$ layer in ZnCoS$_2$.  As  the electronic physics in CuInCo$_2$A$_4$ is also controlled by the CoA$_2$ layer,  their electronic physics  should be described by  the same model derived in ref.\cite{Hu2017ASuperconductors}. However, in CuInCo$_2$A$_4$, the unit cell is doubled at least.  The doubling of the unit cell stems from the CuIn layer due to the inequivalence of Cu and In atoms. We will show that  the effect  of this out-of plane modification  on  the electronic structure of  the CoA$_4$ layer in the first order approximation can be ignored.

\begin{figure}
\centerline{\includegraphics[width=0.5\textwidth]{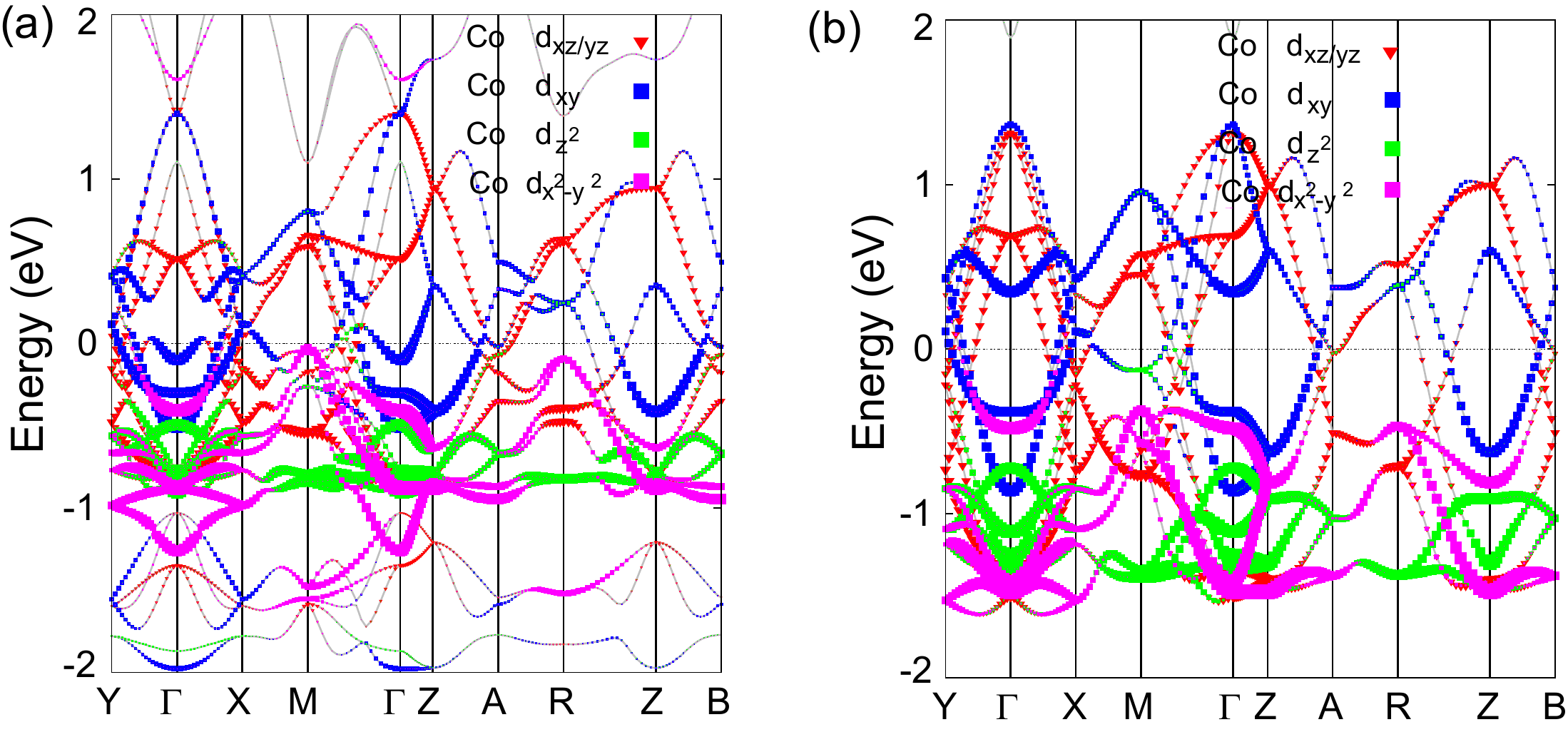}}
\caption{(color online) (a) and (b) are band structures of CuInCo$_2$Te$_4$ and ZnCoS$_2$. The orbital contributions of the different FS sheets are shown color coded: dxz (green online),dyz (red online) and dxy (blue online).
 }
\label{band}
\end{figure}

 In Fig.\ref{band} (a), we plot the band structure of the stannite CuInCo$_2$Te$_4$.  In the figure, the different colors mark the orbital characters.  The three $t_{2g}$ orbitals are  close to half filling and dominate the electronic physics near Fermi energy.  In Fig.\ref{band}(b), we plot the electronic band structure of ZnCoS$_2$ calculated in ref.\cite{Hu2017ASuperconductors} by artificially  doubling the one Co unit cell.   By comparing the Fig.\ref{band} (a) and (b), it is clear that near Fermi energy,  the band structure of CuInCo$_2$Te$_4$ is qualitatively identical to that of ZnCoS$_2$.

%\begin{table}[bt]%The best place to locate the table environment is directly after its first reference in text
%\caption{\label{hopping1} The NN hopping parameters (in unit of eV). x(y) labels the hopping between two NN sites along x(y) directions.}
%\begin{ruledtabular}
%\begin{tabular}{ccccc}
%%\textrm{Left\footnote{Note a.}}& \textrm{Centered\footnote{Note
%%b.}}& \multicolumn{1}{c}{\textrm{Decimal}}&
%%\textrm{Right}\\
%  & $t^{11}_{x}$ & $t^{11}_{y}$ & $t^{33}_{x}$ &band widths $d_{xz/yz}$/$d_{xy}$(eV)\\
% \colrule
%CuInCo$_2$Te$_4$   & 0.52 & 0.03 & 0.24 &   2.18/1.91\\
%CuInCo$_2$Se$_4$ & 0.51 & 0.03 & 0.22    &    2.15/1.73\\
%CuInCo$_2$S$_4$ &0.59 & 0.06 & 0.28    & 2.60/2.24\\
%ZnCoS$_2$\cite{Hu2017ASuperconductors}& 0.63 & 0.07&0.34 & 2.82/2.70
%\end{tabular}
%\end{ruledtabular}
%\label{hopping}
%\end{table}

\begin{table}[bt]%The best place to locate the table environment is directly after its first reference in text
\caption{\label{hopping1} The NN hopping parameters (in unit of eV). x(y) labels the hopping between two NN sites along x(y) directions.}
\begin{ruledtabular}
\begin{tabular}{ccccc}
%\textrm{Left\footnote{Note a.}}& \textrm{Centered\footnote{Note
%b.}}& \multicolumn{1}{c}{\textrm{Decimal}}&
%\textrm{Right}\\
  & $t^{11}_{x}$ & $t^{11}_{y}$ & $t^{33}_{x}$ &band widths $d_{xz/yz}$/$d_{xy}$(eV)\\
 \colrule
CuInCo$_2$Te$_4$   & 0.36 & 0.06 & 0.13 &   2.18/1.91\\
CuInCo$_2$Se$_4$ & 0.35 & 0.06 & 0.12    &    2.15/1.73\\
CuInCo$_2$S$_4$ &0.41 & 0.12 & 0.15    & 2.60/2.24\\
ZnCoS$_2$\cite{Hu2017ASuperconductors}& 0.44 & 0.14&0.18 & 2.82/2.70
\end{tabular}
\end{ruledtabular}
\label{hopping}
\end{table}

We recall the  minimum effective tight binding model Hamiltonian $H_0$ to capture the three $t_{2g}$  orbitals near Fermi surfaces in ref.\cite{Hu2017ASuperconductors}. In the basis of d$_{xz}$, d$_{yz}$, d$_{xy}$  three $t_{2g}$ orbitals, the elements of the $3\times3$ $H_0$ matrix is specified as
\begin{eqnarray}
\label{eq2}
H_{11} & = & \epsilon_{1}+2t^{11}_{x}cos(k_x) +2t^{11}_{y}cos(k_y)+4t^{11}_{xy}cos(k_x)cos(k_y)\nonumber
\\
&& +2t^{11}_{xx}cos(2k_x) +2t^{11}_{yy}cos(2k_y),\nonumber \\
H_{12} & = & -4t^{12}_{xy}sin(k_x)sin(k_y)\nonumber
\\
H_{13} & = & 2it^{13}_xsin(k_x)+4it^{13}_{xy}sin(k_x)cos(k_y)+2it^{13}_{xx}sin(2k_x)\nonumber
\\
H_{22} & = & \epsilon_{2}+2t^{22}_{x}cos(k_x) +2t^{22}_{y}cos(k_y)+4t^{22}_{xy}cos(k_x)cos(k_y)\nonumber
\\
&& +2t^{22}_{xx}cos(2k_x) +2t^{22}_{yy}cos(2k_y),\nonumber
\\
H_{23} & = & 2it^{23}_ysin(k_y)+4it^{23}_{xy}sin(k_y)cos(k_x)+2it^{23}_{xx}sin(2k_y)
\nonumber
\\
H_{33} & = & \epsilon_{3}+2t^{33}_{x}(cos(k_x)+cos(k_y))+4t^{33}_{xy}cos(k_x)cos(k_y)\nonumber
\\
&& +2t^{33}_{xx}(cos(2k_x)+cos(2k_y)),
\end{eqnarray}
where the hopping parameters are specified in ref.\cite{Hu2017ASuperconductors}. In the first order approximation, the electronic band structures of CuInCo$_2$A$_4$ can also be described by this model.   Rather than making a detailed fitting to this model and extracting all parameters for  CuInCo$_2$A$_4$,  we list the most important intra-orbital NN hopping parameters  and the orbital band widths in the Table.\ref{hopping}.  The NN hopping parameters  are slightly larger than those in ZnCoS$_2$\cite{Hu2017ASuperconductors}.

\begin{figure}
\centerline{\includegraphics[width=0.5\textwidth]{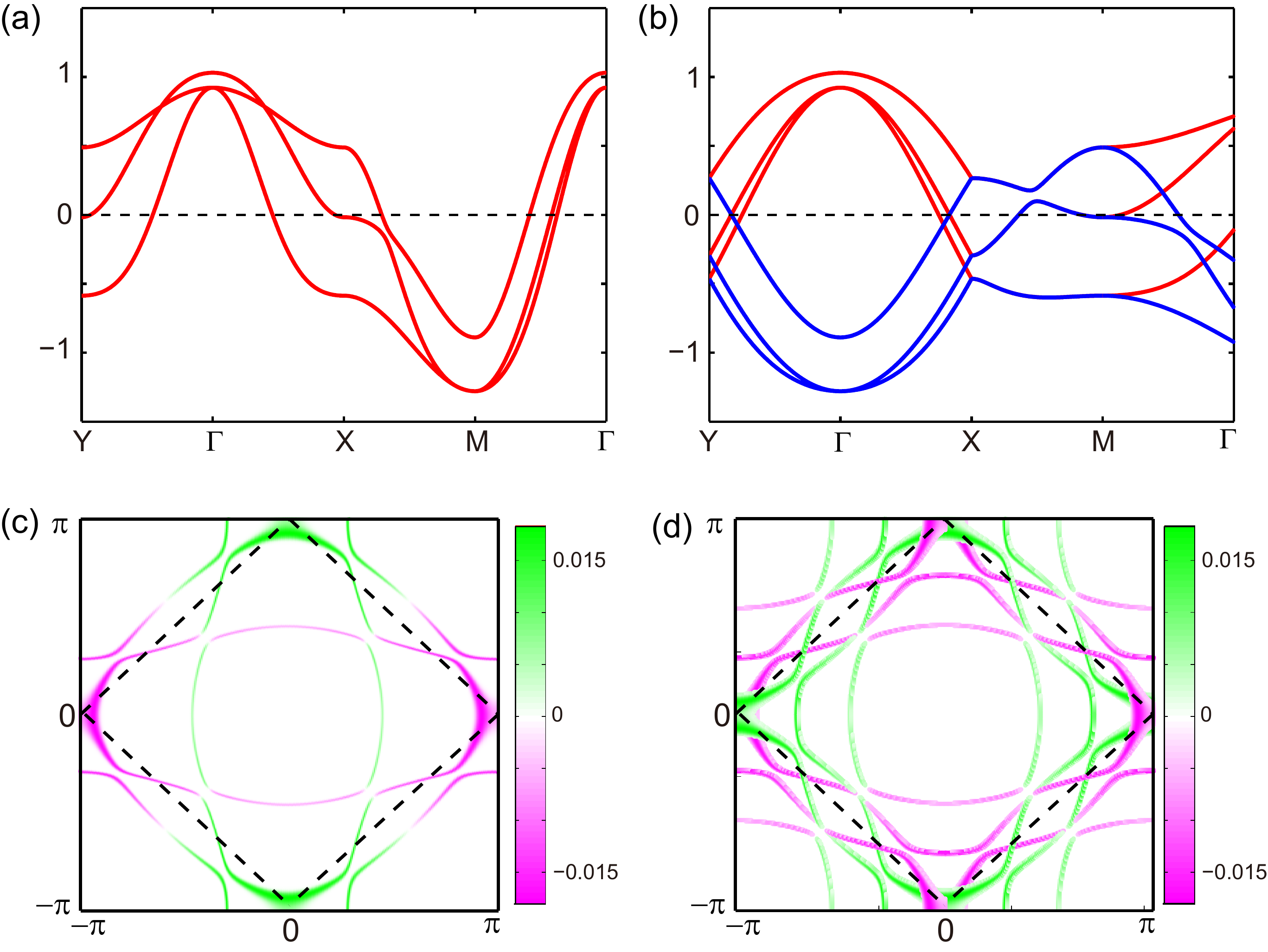}}
\caption{(color online) (a) The unfold band structure and (b) fold band structure of the effective model based on the three d-orbitals. (c) The unfold superconducting gap structure and (d) fold superconducting gap structure of  the d-wave at half filling with $\Delta_0=0.03$ .
 }
\label{gap}
\end{figure}

In ref.\cite{Hu2017ASuperconductors,Li2017,Huding2012}, we have argued  that it is inevitable that the superconducting pairing symmetry in this model  is  a d-wave in which there should be gapless nodes along the diagonal direction in  the momentum space.  To obtain a qualitative understanding of the d-wave sate, we  can make a rough estimation by fixing  the relative intra-orbital pairing strength for three orbitals  according to their hopping energy scales.  In Fig. \ref{gap}, we  draw the three orbital band structure of the effective model in both folded and unfolded  Brillouin zones and  sketch  the d-wave gap form factors in the momentum space by taking $ \frac{1}{3}\Delta_0(cos(k_x)- cos(k_y))$  for $d_{xy}$   and  $\Delta_0cos(k_y)$ for $d_{yz}$, $- \Delta_0 cos(k_x)$ for  $d_{xz}$, where $\Delta_0$ is  a superconducting gap parameter. This state is simply a multi-orbital version of the d-wave state in cuprates.

\section{Discussion}
We have identified a  family of Co-based chalcogenides as potential high-$T_c$ superconductors.  The parental compounds of these materials are multi-orbital Mott insulators with AFM ground states.  As we have calculated theoretically in ref.\cite{Hu2017ASuperconductors},  a  d-wave pairing symmetry is favored in these materials upon doping.  Verifying these predictions does not only add a new family of high-$T_c$ superconductors, but also provides a clear picture and a roadmap to settle illusive high-$T_c$ mechanism.

The Co-based chalcogenides can serve as a bridge to unify cuprates and iron-based superconductors.  On one side, the materials share many similarities with iron-based superconductors.   The electronic physics in both materials are carried out by  $t_{2g}$ multi-orbitals. The Co square lattice in the former is simply a sublattice of the Fe square lattice of  the latter. They both  share  common AFM  interactions.  As we have pointed in this paper,  the trend of the magnetism as a function of transition metal elements  is also very similar.  On the other side,  the materials  share a  common AFM ground  state as  cuprates and have strong Mott physics as well. Both materials are expected to have a common d-wave superconducting state upon doping.

We have mentioned in the introduction that the structure of the Co-based chalcogenides can be considered as a derivative of the insulating zinc-blende chalcogenides by replacing the entire Zn  layer by Co completely.  Electronically, this replacement inserts the d-orbitals into the insulating gap, which is also the case for iron-based superconductors, for example, BaFe$_2$As$_2$ can be considered as  a derivative of  the insulating BaZn$_2$P$_2$ by replacing Zn by Fe entirely\cite{Kamihara2008-jacs}.  This simple understanding can be used to design and find other possible high-$T_c$ materials.

The abrupt quenching of magnetism from the Co to Ni compounds is a natural characteristics of the electronic structure required by the proposed gene condition\cite{Hu-tbp,Hu-genes}. Both cuprates and iron-based superconductors  clearly display the same quenching phenomena. The phenomena can be understood as follows. In the proposed gene condition,  as all d-orbitals around Fermi surfaces are near half-filling and have been  delocalized  through the large d-p hybridization, adding an electron to the d-shell  drastically weakens the electron-electron correlation effect to quench magnetism.

Our theoretical results on the magnetism of the Co  and Ni compounds appear to be consistent with experimental results\cite{Gallardo2016SYNTHESIS4}. In the measured curve of spin susceptibility on the Co-based compounds\cite{Gallardo2016SYNTHESIS4} , the susceptibility increases as temperature increases in the measured temperature window up to 400K.  This indicates that the material is in an AFM state with a Neel transition temperature, $T_N$, higher than 400K which is consistent with our calculated  AFM coupling strength.
The measured curve for the Ni-based compounds does not exhibit such an AFM behavior\cite{Gallardo2016SYNTHESIS4}.

Comparing the Co-based compounds to the previous theoretical compound, ZnCoS$_2$\cite{Hu2017ASuperconductors}, the CoSe$_2$ layer is slightly distorted due to the doubling of the unit cell.  The distortion can affect superconducting transition temperatures $T_c$ as the buckling of CuO layer in cuprates is known to be a strong factor to affect $T_c$\cite{Eisaki2014}. Here the effect could be even larger because the unit cell doubling in a d-wave pairing state can cause the mixture between two bands with opposite pairing signs, which is a destructive reconstruction for pairing.  Therefore, for this family of Co-based materials,  the material quality  can be key  factor in obtaining high-$T_c$. Nevertheless, from T$_N>400$K, the energy scale of the AFM interactions is quite large. We do expect the maximum $T_c$ in this compound should  exceed those of iron-based superconductors.

The electron doping can be achieved by  substituting Co with Ni, just like the substitution of Fe by Co or Ni in iron-based superconductors. The carriers can also be introduced by modifying CuIn layers. In principle, we may consider a modified formula as \ce{Cu_{1-x}In_{1+x}}, \ce{CuIn_{1-x}Sn_{x}} or \ce{CuIn_{1-x}Sr_{x}} to introduce electron carriers or hole carriers.

In summary, we identify  a new family of Co-based high temperature superconductors with a diamond-like stannite or PMCA structure.  The parental materials have been already synthesized  and the measured magnetic properties  are consistent with our predictions.   The materials can serve a bridge to unify the two known high-$T_c$ superconductors, cuprates and iron-based superconductors,  and  provide us a first falsifiable test to the gene condition proposed for unconventional high-$T_c$\cite{Hu-genes} to settle the elusive high-$T_c$ mechanism.

{\it Acknowledgement:}  the work is supported by the Ministry of Science and Technology of China 973 program(Grant No. 2015CB921300, No.~2017YFA0303100), National Science Foundation of China (Grant No. NSFC-11334012), and   the Strategic Priority Research Program of  CAS (Grant No. XDB07000000).

% compare with SR2RUO4
% possible d+is
%lattice distortion and nematicity
%angle dependence, pressure sensitivity
%materials consideration
%\bibliography{hexagonalSC-new,magnetic,M2X2O-layer}

%\bibliography{Mendeley}

%\bibliography{genes}

%\begin{thebibliography}{K2Cr3As3_pairing}
%%\bibitem{Hu2012} J. P. Hu and H. Ding, Sci. Rep. {\bf 2}, 381 (2012).
%%%DFT
%\bibitem{Kresse1993} G. Kresse and J. Hafner, Phys. Rev. B {\bf 47}, 558 (1993).
%\bibitem{Kresse1996} G. Kresse and J. Furthmuller, Comput. Mater. Sci. {\bf 6}, 15 (1996).
%\bibitem{Kresse1996B} G. Kresse and J. Furthmuller, Phys. Rev. B {\bf 54}, 11169 (1996).
%\bibitem{Perdew1996} J. P. Perdew, K. Burke, and M. Ernzerhof, Phys. Rev. Lett. {\bf 77},
%3865 (1996).
%
%\end{thebibliography}

%\begin{thebibliography}
%\bibitem
%\end{thebibliography}

\end{document}